\documentclass[aps,prb,onecolumn,eqsecnum,superscriptaddress]{revtex4-1}
\usepackage{makeidx}
\usepackage{eurosym}
\usepackage{amssymb}
\usepackage{amsfonts}
\usepackage{amsmath}
\usepackage{setspace}
\usepackage[english]{babel}
\usepackage{graphicx}
\usepackage[applemac]{inputenc}
\usepackage{color}
\usepackage{ifthen}
\usepackage{float}
\usepackage{verbatim}
\usepackage{subfigure}
\usepackage[colorlinks,linkcolor=blue,anchorcolor=blue,citecolor=blue,urlcolor=blue]{hyperref}
\usepackage{bm}

\begin{document}
\title{Directional router and controllable non-reciprocity transmission based on phase and pathway coherence}

\author{Xu Yang}
\affiliation{Key Laboratory of Quantum Theory and Applications of MoE, Lanzhou University, Lanzhou, Gansu 730000, China}
\author{Lei Tan}
\email{tanlei@lzu.edu.cn}
\affiliation{Key Laboratory of Quantum Theory and Applications of MoE, Lanzhou University, Lanzhou, Gansu 730000, China}
\affiliation{Key Laboratory for Magnetism and Magnetic Materials of the Ministry of Education, Lanzhou University, Lanzhou $730000$, China}
\author{Wu-Ming Liu}
\email{wliu@iphy.ac.cn}
\affiliation{Beijing National Laboratory for Condensed Matter Physics, Institute of Physics, Chinese Academy of Sciences, Beijing $100190$, China}
\date{\today}

\begin{abstract}
  A multi-channel quantum router with four nodal cavities is constructed by two coupled-resonator waveguides and four single cavities. We can achieve directional routing by adjusting the probability of photon exiting from the specified port to close to 100$\%$ based on multiple pathways between the photon from the incident port to the outgoing port in this hybrid system. Under the effect of phase difference between two classical light fields, the mutual interference between different pathways can be adjusted to destructive interference or constructive interference, which lays the foundation for the increase and decrease of the routing probability. The influence of different parameter values on single photon routing probability is also studied. By studying the analytic formula of probability amplitude, we get the physical mechanism of exiting ports being closed under certain parameter conditions and the phase relationship between the backward transmission and the original direction transmission of photons. Furthermore the non-reciprocal transmission and directional routing beyond chiral coupling can also be realized, which provides new possibilities for the study of quantum routers and new insights for the study of photon transmission characteristics.

\end{abstract}
\maketitle

\section{Introduction}

In recent years, due to its potential applications in quantum information [\onlinecite{aaaa, PhysRevLett.100.170501, PhysRevA.99.022335, Zhang2019, Bao_2023, 10.1088/1674-1056/acc3f6}] processing and quantum networks [\onlinecite{Kimble2008, RevModPhys.87.1379, Pirandola2019, PhysRevLett.125.260506, PhysRevLett.78.3221}], the research on controlling single photon transport through waveguide coupled to quantum emitters is very popular [\onlinecite{RevModPhys.91.015006, Chang2007, Kannan2020, doi:10.1126/science.1155441, Shen:05, PhysRevLett.95.213001, PhysRevLett.101.100501, Witthaut2010, Li2017, PhysRevA.102.053702, Yang_2020, PhysRevA.104.023712, PhysRevLett.126.043602}]. The waveguide can be used as a quantum channel, and the quantum emitter can be placed in the cavity to form a quantum node, which provides a platform for quantum information and quantum computing. As for the carrier of information transmission, photon has become the choice of researchers due to its advantages of fast speed, good robustness and easy control. Since each quantum node is connected to multiple quantum channels, distributing filtered information to different quantum channels is an important problem in the construction of quantum networks. Based on this, the concept of quantum router [\onlinecite{Yan2016}, \onlinecite{PhysRevLett.111.103604}] comes into being. Quantum routers can filter quantum information so that carriers carrying different information enter a specific quantum channel from one quantum channel. Many theoretical researches and experiments on quantum routers have been realized, such as all-optical systems [\onlinecite{
doi:10.1126/science.1254699}, \onlinecite{PhysRevX.3.031013}], cavity QED systems [\onlinecite{PhysRevLett.102.083601, Vahala2003,  doi:10.1126/science.aaj2118}], and photomechanical systems [\onlinecite{PhysRevA.85.021801}, \onlinecite{Ma2017}]. In the study of pure optical systems, the researchers have studied the waveguides with linear dispersion and the coupled-resonator waveguide [\onlinecite{Zang2010}, \onlinecite{PhysRevA.80.033823}] with nonlinear dispersion respectively. A coupled-resonator waveguide [\onlinecite{PhysRevA.93.032337}] is formed by placing a set of cavities with lower decay rate end to end, and coupling the cavity walls between two adjacent cavities. Due to the coupling of cavity walls, photons can be hopped between adjacent cavities and transported in the whole coupled-resonator waveguide. In addition, atoms can be placed in the coupled-resonator waveguide, and the transported photons can be controlled through the coupling of atoms and photons. Because the dispersion relation of coupled-resonator waveguide is nonlinear, the hybrid system composed of coupled-resonator waveguide has wider bandwidth and stronger interaction between light and matter, and some novel phenomena may be produced. As a basic quantum device, coupled-resonator waveguide becomes a reliable platform for studying the interaction between light and matter.

After quantum routers can distribute photons into different channels, directional routing and non-reciprocal routing become the focus of research. Directional routing can output the required information more accurately and quickly, while non-reciprocal transmission can be used as a means of information filtering, closing the bidirectional transmission of specific information, so that it can only be one-way transmission. In general, single-photon quantum router is mainly composed of two coupled-resonator waveguide and a single quantum transmitter [\onlinecite{Wang2021}]. In the multi-channel quantum router model of this configuration, incident photons are output symmetrically at four ports, and the maximum routing probability is only 50$\%$ [\onlinecite{Du2021}]. This feature limits the application of quantum routers to some extent. Therefore, it is still a very important problem to study the full transfer rate is 100$\%$, that is, directional routing in quantum routers. Unlike quantum routers coupling single quantum transmitters, the coupling of multiple quantum transmitters with multiple one-dimensional coupled-resonator waveguide can provide many novel phenomena and contribute to the further application of quantum routers [\onlinecite{PhysRevA.98.063809}, \onlinecite{Yan2021}]. In the previous study of non-reciprocal transmission, it was realized through the chiral coupling between the waveguide and the transmitter [\onlinecite{PhysRevA.105.013711}, \onlinecite{PhysRevLett.126.043602}]. However, chiral coupling requires the emitter to be coupled only to photons transported in a single direction in the channel. Therefore, we hope that the non-reciprocal device without chiral coupling can be realized by adjusting the coupling strength and the parameters of the classical field [\onlinecite{Wang_2022}], and the non-reciprocal direction can be changed by changing the parameters. Compared with the previous non-reciprocal transmission, the change of the direction of non-reciprocal transmission can make the method of information transmission more flexible. By increasing the number of nodal cavities, we can increase the number of photon transport pathways in the router. Different pathways interfere with each other to realize that the probability of photon exiting from one or some ports under certain parameter conditions is cancelled by interference. In order to increase the controllability of the model, two-level atoms are placed in two nodal cavities, and two $\Delta$-type three-level atoms [\onlinecite{Xiao2020}, \onlinecite{PhysRevA.97.023821}] are placed in the other two nodal cavities, and classical light fields of different phases are coupled to the transition between two excited states of atoms. We focus on the effect of phase difference and coupling strength on the single photon transport characteristics.

This paper is organized as follows: In Sec. $\rm{\uppercase\expandafter{\romannumeral2}}$, the model is introduced. In Sec. $\rm{\uppercase\expandafter{\romannumeral3}}$, we study the scattering process of single-photon by using a discrete-coordinate scattering approach and get the expressions of three amplitudes. We will study the effect of position, phase difference, Rabi frequencies and coupling strengths, and discuss the underlying physics for controlling single-photon routing in Sec. $\rm{\uppercase\expandafter{\romannumeral4}}$. In Sec. $\rm{\uppercase\expandafter{\romannumeral5}}$, we investigate the possibility of realizing non-reciprocal transport in this model, and change the non-reciprocal direction by regulating the phase. Finally, the results are concluded in Sec. $\rm{\uppercase\expandafter{\romannumeral6}}$.

\section{Model and Hamiltonian}

\begin{figure}
  \centering
  \includegraphics[width=14cm,height=8cm]{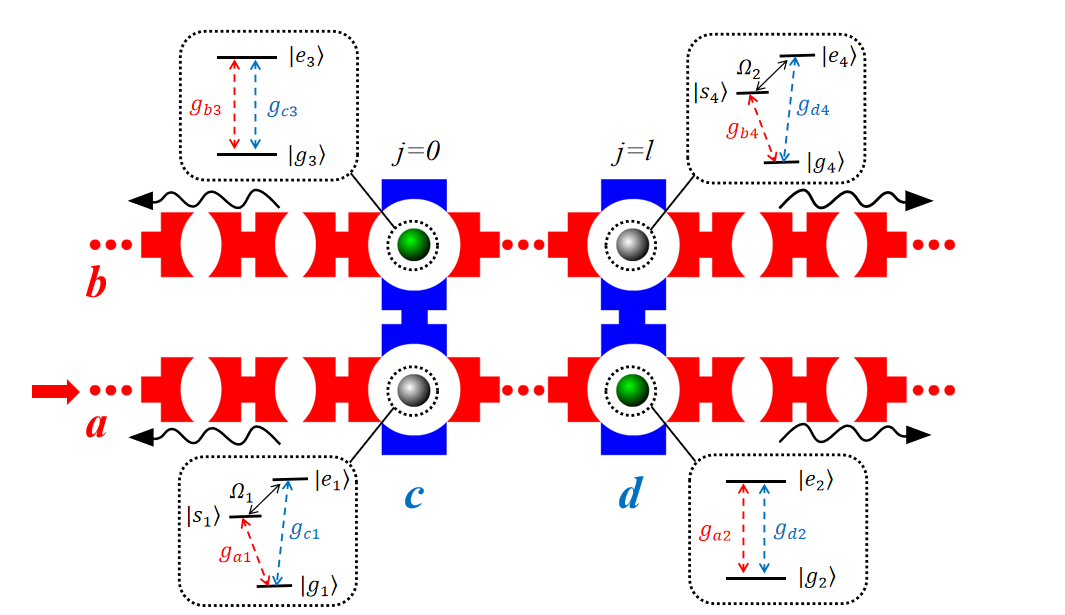}\\
  \caption{(Color online) Schematic routing of single photons in two parallel CRWs. The connection between the two CRWS is through the blue vertical cavity. Four atoms are placed in four intersecting cavities, which we refer to as nodes. The blue dashed line represents the transition of the atoms coupled to the cavity mode of the vertical cavity, the red dashed line represents the transition coupled to the cavity mode of the horizontal cavities. The thin black solid line indicates that the atom couples the classical field and transitions between the excited state and intermediate state. The four wavy lines represent photons exiting from the four ports and the probabilities are $L^{a}$, $R^{a}$, $L^{b}$ and $R^{b}$. }\label{Fig1}
\end{figure}

As shown in Fig. 1, the considered hybrid system is composed of two horizontal one-dimensional coupled-resonator waveguides (CRWs) and four vertical cavities. The upper CRW is CRW-b and the lower one is CRW-a. The vertical cavities at $j=0$ is named Cavity-C, and these at $j=l$ is named Cavity-D, where $j$ is the position of the cavity of two CRWs with $j=-\infty,...,\infty$. The cavity modes are described by the creation (annihilation) operator $x^{\dag}_{j}(x_j)$ in CRWs and $y^{\dag}_{x}(y_{x})$ in vertical cavities, where $x=a,b$ and $y=c,d$. For simplicity, we assume that the hopping coefficient between the nearest-neighbor cavities is the same.

Four vertical cavities and CRWs form four nodes embedded with an atom. Two two-level atoms with an excited state $|e_{m}\rangle$ and a ground state $|g_{m}\rangle$ are placed in a node composed of CRW-a and Cavity-$D_{a}$ (named Atom-2) and a node composed of CRW-b and Cavity-$C_{b}$ (named Atom-3). Two $\Delta$-type three-level atoms are placed in a node composed of CRW-a and Cavity-$C_{a}$ (named Atom-1) and a node composed of CRW-b and Cavity-$D_{b}$ (named Atom-4), respectively. The $\Delta$-type three-level atom characterized by a ground state $|g_{m}\rangle$, an intermediate state $|s_{m}\rangle$ and an excited state $|e_{m}\rangle$, where $m$ is the number of the atom. For the $\Delta$-type atoms, a classical light field with frequency $\nu$ couples to the $|e_{m}\rangle \leftrightarrow |s_{m}\rangle$ transition with Rabi frequency $\Omega$. The cavity field of CRWs (vertical cavities) couples to the transition with strength $g_{xm}(g_{ym})$.

The total Hamiltonian of the hybrid system reads ($\hbar=1$)
\begin{equation}
\begin{split}
H&=\sum_{j}[\omega_{a}a^{\dag}_{j}a_{j}-\xi(a^{\dag}_{j+1}a_{j}+H.c.)+\omega_{b}b^{\dag}_{j}b_{j}-\xi(b^{\dag}_{j+1}b_{j}+H.c.)]\\
&+\omega_{c}(c^{\dag}_{a}c_{a}+c^{\dag}_{b}c_{b})-\xi(c^{\dag}_{a}c_{b}+H.c.)+\omega_{d}(d^{\dag}_{a}d_{a}+d^{\dag}_{b}d_{b})-\xi(d^{\dag}_{a}d_{b}+H.c.)\\
&+\omega_{s1}|s_{1}\rangle\langle s_{1}|+\omega_{e1}|e_{1}\rangle\langle e_{1}|+\omega_{e2}|e_{2}\rangle\langle e_{3}|+\omega_{e3}|e_{3}\rangle\langle e_{3}|+\omega_{e4}|e_{4}\rangle\langle e_{4}|+\omega_{s4}|s_{4}\rangle\langle s_{4}|\\
&+g_{a1}a^{\dag}_{0}|s_{1}\rangle\langle g_{1}|+g_{a2}a^{\dag}_{l}|e_{2}\rangle\langle g_{2}|+g_{b3}b^{\dag}_{0}|e_{3}\rangle\langle g_{3}|+g_{b4}b^{\dag}_{l}|s_{4}\rangle\langle g_{4}|+H.c.\\
&+g_{c1}c^{\dag}_{a}|e_{1}\rangle\langle g_{1}|+g_{c3}c^{\dag}_{b}|e_{3}\rangle\langle g_{3}|+g_{d2}d^{\dag}_{a}|e_{2}\rangle\langle g_{2}|+g_{d4}d^{\dag}_{b}|e_{4}\rangle\langle g_{4}|+H.c.\\
&+\Omega_{1}|e_{1}\rangle\langle s_{1}|e^{-i(\nu t+\phi)}+\Omega_{2}|e_{4}\rangle\langle s_{4}|e^{-i\nu t}+H.c.,
\end{split}
\label{eq1}
\tag{1}
\end{equation}
where $\omega_a$, $\omega_{b}$, $\omega_{c}$ and $\omega_{d}$ are the frequency of the cavity mode $a_{0}$, $b_{0}$, $c_{a}$ and $d_{a}$, respectively. All the cavities are considered to have the same frequency in CRW-a (CRW-b) for simplicity and the frequencies of four vertical cavities are also the same. $\xi$ is hopping strength of any two neighbour cavities, $a^{\dag}_{j+1}a_{j}(b^{\dag}_{j+1}b_{j})$ means to annihilate a photon in the $j$th cavity and create another one in the $(j+1)$th cavity in two CRWs. $\omega_{e1}$, $\omega_{e2}$, $\omega_{e3}$ and $\omega_{e4}$ are the frequencies of energy levels $|e\rangle$ in four atoms. $\omega_{s1}$ and $\omega_{s4}$ are the frequencies of $|s\rangle$ in Atom-1 and Atom-4, and the energy of the ground state $|g\rangle$ is chosen as zero point. $g_{xm}(g_{ym})$  is the atom-photon coupling strength. $\phi$ is the phase difference between two classical light fields. If the photon couples Atom-1 and Atom-4 respectively in different pathways, $\phi$ can be regarded as the phase difference between two pathways. To remove the time-dependent factor of the Hamiltonian, we rewrite the Hamiltonian in a rotating frame of reference based on the unitary transformation $U=exp[i\nu t(|s_{1}\rangle\langle s_{1}|+|s_{4}\rangle\langle s_{4}|)]$. The Hamiltonian under rotating frame is $H'=U^{\dag}HU-iU^{\dag}\partial_{t}U$. The total Hamiltonian $H$ becomes
\begin{equation}
\begin{split}
H'&=\sum_{j}[\omega_{a}a^{\dag}_{j}a_{j}-\xi(a^{\dag}_{j+1}a_{j}+H.c.)+\omega_{b}b^{\dag}_{j}b_{j}-\xi(b^{\dag}_{j+1}b_{j}+H.c.)]\\
&+\omega_{c}(c^{\dag}_{a}c_{a}+c^{\dag}_{b}c_{b})-\xi(c^{\dag}_{a}c_{b}+H.c.)+\omega_{d}(d^{\dag}_{a}d_{a}+d^{\dag}_{b}d_{b})-\xi(d^{\dag}_{a}d_{b}+H.c.)\\
&+\omega'_{s1}|s_{1}\rangle\langle s_{1}|+\omega_{e1}|e_{1}\rangle\langle e_{1}|+\omega_{e2}|e_{2}\rangle\langle e_{3}|+\omega_{e3}|e_{3}\rangle\langle e_{3}|+\omega_{e4}|e_{4}\rangle\langle e_{4}|+\omega'_{s4}|s_{4}\rangle\langle s_{4}|\\
&+g_{a1}a^{\dag}_{0}|s_{1}\rangle\langle g_{1}|+g_{a2}a^{\dag}_{l}|e_{2}\rangle\langle g_{2}|+g_{b3}b^{\dag}_{0}|e_{3}\rangle\langle g_{3}|+g_{b4}b^{\dag}_{l}|s_{4}\rangle\langle g_{4}|+H.c.\\
&+g_{c1}c^{\dag}_{a}|e_{1}\rangle\langle g_{1}|+g_{c3}c^{\dag}_{b}|e_{3}\rangle\langle g_{3}|+g_{d2}d^{\dag}_{a}|e_{2}\rangle\langle g_{2}|+g_{d4}d^{\dag}_{b}|e_{4}\rangle\langle g_{4}|+H.c.\\
&+\Omega_{1}|e_{1}\rangle\langle s_{1}|e^{-i\phi}+\Omega_{2}|e_{4}\rangle\langle s_{4}|+H.c.,
\end{split}
\label{eq2}
\tag{2}
\end{equation}
where $\omega'_{s1}=\omega_{s1}+\nu$ and $\omega'_{s4}=\omega_{s4}+\nu$. In the hybrid system, the total excitation operator can be written as
\begin{equation}
N=\sum_{j,x,y}(x^{\dag}_{j}x_{j}+y^{\dag}_{x}y_{x})+|e_{1}\rangle\langle e_{1}|+|e_{2}\rangle\langle e_{2}|+|e_{3}\rangle\langle e_{3}|+|e_{4}\rangle\langle e_{4}|+|s_{1}\rangle\langle s_{1}|+|s_{4}\rangle\langle s_{4}|.
\label{eq3}
\tag{3}
\end{equation}

\section{scattering process of single-photon}
By using Fourier transform $d_{k}=\frac{1}{\sqrt{N}}\sum_{k}exp(ikj)d_{j}(d=a,b)$ and the periodic boundary conditions, one can get
\begin{flalign}
E_{k_{a}}&=\omega_{a}-2\xi{\rm cos}k_{a},
\label{eq4a}
\tag{4a}\\
E_{k_{b}}&=\omega_{b}-2\xi{\rm cos}k_{b}.
\label{eq4b}
\tag{4b}
\end{flalign}
Equations. (4a) and (4b) describe the nonlinear dispersion relation in each bare CRW, indicating that each CRW possesses an energy band with bandwidth $4\xi$.
In the single-excitation subspace, the eigenstate of the system can be written as
\begin{equation}
\begin{split}
|\Psi\rangle=\sum_{j}(\alpha_{j}a_{j}^{\dag}+&\beta_{j}b_{j}^{\dag})|g_{1}g_{2}g_{3}g_{4}\emptyset\rangle+(\mathcal{C}_{a}c^{\dag}_{a}+\mathcal{C}_{b}c^{\dag}_{b}+\mathcal{D}_{a}d^{\dag}_{a}+\mathcal{D}_{b}d^{\dag}_{b})|g_{1}g_{2}g_{3}g_{4}\emptyset\rangle\\
&+v_{1}|s_{1}g_{2}g_{3}g_{4}\emptyset\rangle+v_{2}|e_{1}g_{2}g_{3}g_{4}\emptyset\rangle+v_{3}|g_{1}e_{2}g_{3}g_{4}\emptyset\rangle\\
&+v_{4}|g_{1}g_{2}e_{3}g_{4}\emptyset\rangle+v_{5}|g_{1}g_{2}g_{3}s_{4}\emptyset\rangle+v_{6}|g_{1}g_{2}g_{3}e_{4}\emptyset\rangle,
\end{split}
\label{5}
\tag{5}
\end{equation}
where $|\emptyset\rangle$ is the vacuum state in CRWs and vertical cavity. The first two terms describe the single photon freely propagates in two CRWs, and $\alpha_{j}(\beta_{j})$ are the probability amplitudes of single-photon state in the $j$th cavity in CRW-a (CRW-b). The other four terms in the first line describe the single photon freely propagates in four vertical cavities. The terms in the second line denote the Atom-m populated in its excited state or intermediate state by absorbing a single photon, where $v_1...v_6$ are probability amplitudes of these states, respectively.

From the eigenequation $H'|\Psi\rangle=E|\Psi\rangle$, we can get a series of coupled steady equations for all amplitudes. Removing other amplitudes, we can rewrite discrete scattering equation of single photon as
\begin{flalign}
(E-\omega_{a})\alpha_{j}=-\xi(\alpha_{j+1}+\alpha_{j-1})+\delta_{j0}(\frac{X_{1}}{V_{1}}\alpha_{j}-e^{i\phi}\frac{X_{2}}{V_{1}}\beta_{j})+\delta_{jl}(\frac{X_{3}}{V_{2}}\alpha_{j}-\frac{X_{4}}{V_{2}}\beta_{j}),
\label{eq6a}
\tag{6a}\\
(E-\omega_{b})\beta_{j}=-\xi(\beta_{j+1}+\beta_{j-1})-\delta_{j0}(e^{-i\phi}\frac{X_{2}}{V_{1}}\alpha_{j}-\frac{X_{5}}{V_{1}}\beta_{j})-\delta_{jl}(\frac{X_{4}}{V_{2}}\alpha_{l}-\frac{X_{1}}{V_{2}}\beta_{j}).
\label{eq6b}
\tag{6b}
\end{flalign}
For simplicity, we set the frequencies of all excited and intermediate states to be the same as those of the cavity modes, $\omega_{x}=\omega_{y}=\omega_{em}=\omega'_{sm}=\omega$. Coupling strengths is divided into two parts, one is the coupling strengths between CRWs and atoms, which is $g_{xm}=g_{a}$, and the other is the coupling strengths between vertical cavities and atoms which is represented by $g_{ym}=g_{s}$. Then the parameters can be expressed as
\begin{flalign}
&X_{1}=g_{a}^{2}(-g_{s}^{2}+(E-\omega)(E+\xi-\omega))(E^{2}-g_{s}^{2}+\omega(\xi+\omega)-E(\xi+2\omega)),
\label{eq7a}
\tag{7a}\\
&X_{2}=g_{a}^{2}g_{s}^{2}\Omega_{1}\xi,
\label{eq7b}
\tag{7b}\\
&X_{3}=g_{a}^{2}(E^{4}+\Omega_{2}^{2}\xi^{2}-4E^{3}\omega-(g_{s}^{2}+\Omega_{2}^{2}+\xi^{2})\omega^{2}+\omega^{4}-E^{2}(g_{s}^{2}+\Omega_{2}^{2}+\xi^{2}-6\omega^{2})+2E\omega(g_{s}^{2}+\Omega_{2}^{2}+\xi^{2}-2\omega^{2})),
\label{eq7c}
\tag{7c}\\
&X_{4}=g_{a}^{2}g_{s}^{2}\Omega_{2}\xi,
\label{eq7d}
\tag{7d}\\
&X_{5}=g_{a}^{2} (E^{4} + \Omega_{1}^{2} \xi^{2} - 4 E^{3} \omega - (g_{s}^{2} + \Omega_{1}^{2} + \xi^{2}) \omega^{2} + \omega^{4} -E^{2} (g_{s}^{2} + \Omega_{1}^{2} + \xi^{2} - 6 \omega^{2}) +2 E \omega (g_{s}^{2} + \Omega_{1}^{2} + \xi^{2} - 2 \omega^{2})),
\label{eq7e}
\tag{7e}
\end{flalign}
 and $V_{1(2)}=(E-\omega)(E^{4}+g_{s}^{4}-4E^{3}\omega+(\Omega_{1(2)}^{2}-\omega^{2})(\xi^{2}-\omega^{2})-E^{2}(2g_{s}^{2}+ \Omega_{1(2)}^{2}+\xi^{2}-6\omega^{2})+g_{s}^{2}(\Omega_{1(2)}^{2}-2\omega^{2})+2E\omega(2g_{s}^{2}+\Omega_{1(2)}^{2}+\xi^{2}-2\omega^{2}))$.

We can rewrite $V_{1(2)}$ as
\begin{equation}
V_{1(2)}=(E-\omega_{0})(E-\omega_{+1(2)})(E-\omega_{-1(2)})(E-\omega'_{+1(2)})(E-\omega'_{-1(2)}).
\label{eq8}
\tag{8}
\end{equation}
with
\begin{flalign}
&                  \omega_{0}=\omega,
\label{eq9a}
\tag{9a}\\
&\omega_{-1(2)}=\frac{1}{2}(-\sqrt{2}\sqrt{2g_{s}^{2}+\Omega_{1(2)}^{2}+\xi^{2}-\sqrt{\Omega_{1(2)}^{2}+4g_{s}^{2}\xi^{2}-2\Omega_{1(2)}^{2}\xi^{2}+\xi^{4}}}+2\omega),
\label{eq9b}
\tag{9b}\\
&\omega_{+1(2)}=\frac{1}{2}(\sqrt{2}\sqrt{2g_{s}^{2}+\Omega_{1(2)}^{2}+\xi^{2}-\sqrt{\Omega_{1(2)}^{2}+4g_{s}^{2}\xi^{2}-2\Omega_{1(2)}^{2}\xi^{2}+\xi^{4}}}+2\omega),
\label{eq9c}
\tag{9c}\\
&\omega'_{-1(2)}=\frac{1}{2}(-\sqrt{2}\sqrt{2g_{s}^{2}+\Omega_{1(2)}^{2}+\xi^{2}+\sqrt{\Omega_{1(2)}^{2}+4g_{s}^{2}\xi^{2}-2\Omega_{1(2)}^{2}\xi^{2}+\xi^{4}}}+2\omega),
\label{eq9d}
\tag{9d}\\
&\omega'_{+1(2)}=\frac{1}{2}(\sqrt{2}\sqrt{2g_{s}^{2}+\Omega_{1(2)}^{2}+\xi^{2}+\sqrt{\Omega_{1(2)}^{2}+4g_{s}^{2}\xi^{2}-2\Omega_{1(2)}^{2}\xi^{2}+\xi^{4}}}+2\omega).
\label{eq9e}
\tag{9e}
\end{flalign}
Equations. (9a)-(9e) are defined as the scattering frequencies. The four nodes composing the router and the four atoms placed in the nodes are regarded as a whole, just like the atoms and the classical field in the dressed state. When the frequency $\omega_{in}$ of the incident photon is equal to the above frequencies, it will be absorbed by the router and scattered from the four ports. This can be shown in the following. The wave functions can be given by
\begin{flalign}
\alpha_{j}=&\left\{
\begin{array}{lr}
e^{ik_{a}j}+l^{a}e^{-ik_{a}j},&j<0,\\
r_{1}e^{ik_{a}j}+l_{1}e^{-ik_{a}j},&0<j<l,\\
r^{a}e^{ik_{a}j},&j>l,
\end{array}
\right.
\label{eq10a}
\tag{10a}\\
\beta_{j}=&\left\{
\begin{array}{lr}
l^{b}e^{-ik_{b}j},&j<0,\\
r_{2}e^{ik_{b}j}+l_{2}e^{-ik_{b}j},&0<j<l,\\
r^{b}e^{ik_{b}j},&j>l,
\end{array}
\right.
\label{eq10b}
\tag{10b}
\end{flalign}
where $l^{a}$ $(l^{b})$ is the amplitude of the left-going photon of CRW-a (CRW-b) from the cavity $a_{0}$ $(b_{0})$, and $r^{a}$ $(r^{b})$ is the amplitude of the right-going photon of CRW-a (CRW-b) from the cavity $a_{l}$ $(b_{l})$. Substituting Eqs. (10a) and (10b) into (6a) and (6b), the amplitudes as functions of the incident energy $E$ can be obtained for $E=E_{k_{a}}=E_{k_{b}}$ at far away from $j=0$ and $j=l$ site. The amplitudes are shown in the appendix. One can define the outgoing probability of each port $R^{a}(E)=|r^{a}(E)|^{2}$, $L^{a}(E)=|l^{a}(E)|^{2}$, $R^{b}(E)=|r^{b}(E)|^{2}$ and $L^{b}(E)=|l^{b}(E)|^{2}$, we call them routing rates of four-port. The coefficients in the appendix can be expressed as $o^{x}=C(o^{x})/G$, where $o=l,r$, it can easily get $|C(l^{a})|^{2}+|C(l^{b})|^{2}+|C(r^{a})|^{2}+|C(r^{b})|^{2}=|G|^{2}.$ Therefore we consider $1/G$ to be the normalized factor and $C(o^{x})$ to represent the amplitude before the superposition of all wave functions. Meanwhile, since the photons are incident from the left port of CRW-a, we can also denote the outgoing photons from the left port of CRW-a as reflection and the outgoing photons from the right side of CRW-a as transmission. Note that $R^{a}+R^{b}+L^{a}+L^{b}=1$ physically. The photon routing can be regulated by the coupling strength, Rabi frequencies, phase difference and position. In the following, we will change the parameters mentioned above and study their effects on the photon routing.

\section{The effect of parameters on spectra}

As the position $l$ changes, the number of horizontal empty cavities between the two atoms increases, which leads to some interesting phenomena. For simplicity, we set the Rabi frequencies of atom-1 and atom-4 to be the same, and then we can get $X_{2}=X_{4}$, $X_{3}=X_{5}$ and $V_{1}=V_{2}$. Figure 2(a) and 2(b) show the spectra of four routing rates as a function of incident energy $E$ when $l$ is equal to 6 and 20 respectively. We set $\xi$ as the reduced unit and omit it in the following figure notes. In order to clearly show the difference between the frequency of the incident photon and the frequency of the cavity mode, we replace $E$ with $\Delta+\omega$ in the equations in the appendix and plot the following figures. By comparing Fig. 2(a) and (b), we find that as $l$ increases, the number of peaks in all spectra describing outgoing photons increases. The formula of the form $A_{0}+A_{1}cos(2kl)+A_{2}sin(2kl)+A_{3}cos(4kl)+A_{4}sin(4kl)$ can be obtained by simplifying $|C(o^{x})|^{2}$ in the previous section, where $A_{n}$ are functions of wave vector $k$ and other parameters. From this simplified form, we can find that with the increase of $l$, the oscillation of probability will increase accordingly, which is reflected in the increase of the number of peaks in the spectra. From Eqs. (10a) and (10b), it can be seen that when a photon is excited into a cavity where the position $j$ is greater than 0 and less than $l$, it can be transmitted either to the left or to the right. Thus, there are at least two pathways for the photon to hopping from the cavity at position $j=0$ to the cavity at position $j=l=2$. The two pathways can be expressed as
\begin{flalign}
&Pathway_{1}:x^{\dag}_{0}|\emptyset\rangle\rightarrow x^{\dag}_{1}|\emptyset\rangle\rightarrow x^{\dag}_{2}|\emptyset\rangle,
\tag{P1}\\
&Pathway_{2}:x^{\dag}_{0}|\emptyset\rangle\rightarrow x^{\dag}_{1}|\emptyset\rangle\rightarrow x^{\dag}_{0}|\emptyset\rangle\rightarrow x^{\dag}_{1}|\emptyset\rangle\rightarrow x^{\dag}_{2}|\emptyset\rangle,
\tag{P2}
\end{flalign}
where $x$ can be represented as either $a$ or $b$ as above, and $|\emptyset\rangle$ is the vacuum state in CRWs. These pathways interfere with each other. When $l=3$, the incident photon in both cavities of $j=1$ and $j=2$ can be transported to the left cavity, and there are more pathways from the cavity at position $j=0$ to the cavity of $j=3$ than to $j=2$. So as the number of cavities between the two cavities increases, so does the number of interference pathways that the photon takes to transmit from the origin to the destination. The result shown in the spectra is that the number of peaks increases as $l$ increases.

\begin{figure}[ht]
  \centering
  \begin{subfigure}
  \centering
  \includegraphics[width=7cm,height=5.5cm]{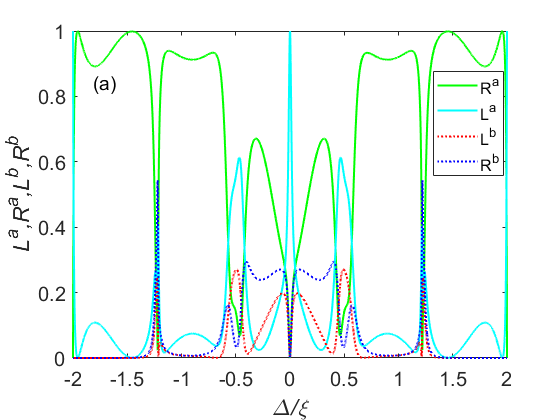}
  \end{subfigure}
    \centering
  \begin{subfigure}
  \centering
  \includegraphics[width=7cm,height=5.5cm]{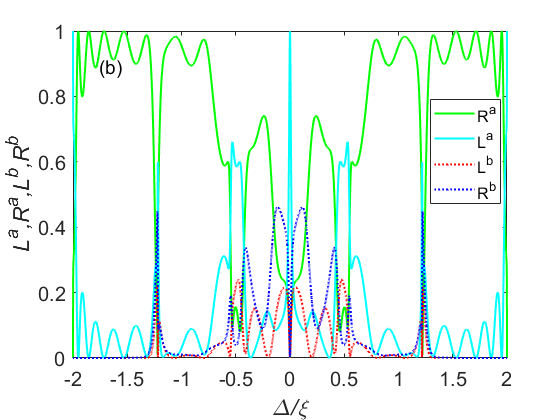}
  \end{subfigure}
    \centering
  \begin{subfigure}
  \centering
  \includegraphics[width=7cm,height=5.5cm]{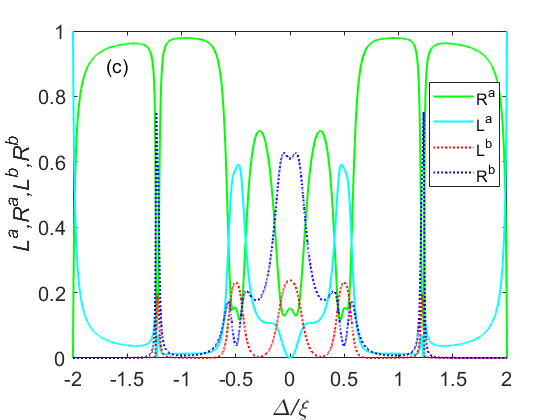}
  \end{subfigure}
    \centering
  \begin{subfigure}
  \centering
  \includegraphics[width=7cm,height=5.5cm]{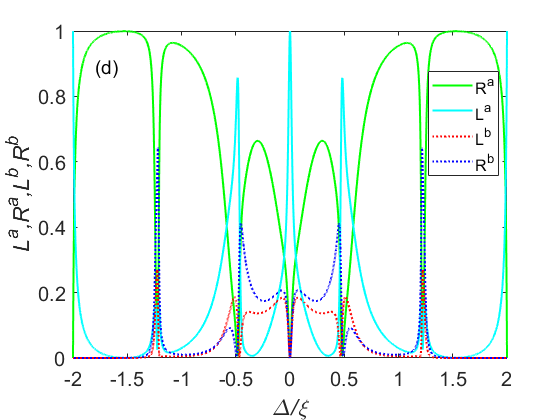}
  \end{subfigure}
  \caption{(Color online) $L^{a}(E)$ (solid cyan line), $L^{b}(E)$ (dashed red line), $R^{a}(E)$ (solid green line) and $R^{b}(E)$ (dashed blue line) as a function of the energy of the incident wave. (a) $l$=6, the fluctuation of the line is weak. (b) $l$=20, the fluctuation of the line is strong. (c) $l$=1, only $L^{a}(E)$ is zero when $\Delta=0$. (d) $l$=2, only $L^{a}$ is 1, and all the other three rate are 0. For convenience, the other parameters are $\xi=1$, $g_{a}=g_{s}=0.5\xi$, $\Omega_{1}=\Omega_{2}=0.5\xi$, $\omega=10\xi$. }
\end{figure}

When $l$ is close, different values of $l$ will also make the spectra very different. Figs. 2(c) and (d) show the spectra of four routing rate as a function of detuning $\Delta$ when $l$ is equal to 8 and 9 respectively. It can be found that when $l$ is even, $R^{a}$ is 1 and the other three routing rates are zero at the resonance energy $\Delta=0$, but when $l$ is odd, all of the four routing rates are nonzero. That is to say, when $l$ is even, the resonant photon incident from the left side of CRW-a is bound to reflect, then the other three output ports are forbidden. When $l$ is odd, the resonant photon may be output from any of the four ports. To explain this phenomenon, we rewrite Eqs. (A2), (A3) and (A4) for the other three amplitudes in the form $\sum_{n}C_{n}(e^{2ikl}-1)^{n}/\sum_{n}C'_{n}(e^{2ikl}-1)^{n}$, but for $l^{a}$, the form is $\sum_{n}C_{n}(e^{2ikl}-1)^{n}/\sum_{n}C'_{n}(e^{2ikl}-1)^{n}-1$, where $n=0,1,2$, $C_{n}$ and $C'_{n}$ are the functions of wave vector $k$. From Eqs. (4a), (4b) and $\Delta=E-\omega$ we can find that when $\Delta=0$, $k$ is equal to $\pi/2$. It is easy to find that $e^{2ikl}-1$ is zero under the condition that $k=\pi/2$ and $l$ is even, in which the values of the three amplitudes except $l^{a}$ are only represented by the expression $C_{0}/C'_{0}$ without $l$ while $r^{a}$ is expressed as $C_{0}/C'_{0}-1$. Meanwhile, by calculation we find that $C_{0}$ is much smaller than $C'_{0}$ as $k$ approaches $\pi/2$. Thus under the condition that $l$ is even, only the amplitude $l^{a}$ is not equal to zero at the resonance frequency and is totally reflected in the spectra.

Next we will investigate the effect of phase on the probability of photon routing at each port. As the phase changes, the routing rate of photons exiting from different ports also has different characteristics. In this case, we still take $\Omega_{1}=\Omega_{2}$. The amplitude of the photon output from each port can be regarded as the superposition of all the coherent paths of the photon transported in the router, which is $C(o ^{x})\Psi(o^{x})=\sum_{n}u_{n}(o^{x})\psi_{n}(o^{x})$, where $o=l,r$ and $n=1,2,...$. $\Psi(o^{x})$ is the total wave function of the photon exits from the $o$-side of CRW-$x$, $\psi_{n}(o^{x})$ is the n-th pathway of photon transport in the router, and $u_{n}(o^{x})$ is the corresponding amplitude of this pathway.

 In order to clearly study the variation of routing rate $R^{b}$ with wave vector and phase, we plot Figs. 3(a) and 3(b). From these two figures we find that the value of $R^{b}$ is 0 when the phase difference is $\pi$ and does not change with the wave vector $k$ and position $l$. This means that when $\Omega_{1}=\Omega_{2}$, we can set the phase difference to $\pi$ and thus turn off the outgoing photon at the right port of the CRW-b. The phenomenon can be analyzed from the scheme in Fig. 1. Since this phenomenon is not affected by position $l$, we next take $l=1$ to simplify the analysis. When the outgoing port of the photon is selected as the right port of the CRW-b, there are two pathways for photon transport, and these two pathways interfere with each other. The two pathways are shown as follows:

\begin{flalign}
\begin{split}
Pathway_{3}:&a^{\dag}_{0}|\emptyset \mathcal{G} \rangle\xrightarrow[a^{\dag}_{l}a_{0}]{} a^{\dag}_{l}|\emptyset \mathcal{G}\rangle \xrightarrow[a_{l}|e_{2}\rangle\langle g_{2}|]{} |e_{2}\rangle\langle g_{2}||\emptyset \mathcal{G}\rangle\xrightarrow[d^{\dag}_{a}|g_{2}\rangle\langle e_{2}|]{} d^{\dag}_{a}|\emptyset \mathcal{G}\rangle\xrightarrow[d^{\dag}_{b}d_{a}]{}\\
&d^{\dag}_{b}|\emptyset \mathcal{G}\rangle\xrightarrow[d_{b}|e_{4}\rangle\langle g_{4}|]{} |e_{4}\rangle\langle g_{4}|\emptyset \mathcal{G}\rangle\xrightarrow[|s_{4}\rangle\langle e_{4}|]{} |s_{4}\rangle\langle g_{4}|\emptyset \mathcal{G}\rangle\xrightarrow[b^{\dag}_{l}|g_{4}\rangle\langle s_{4}|]{} b^{\dag}_{l}|\emptyset \mathcal{G}\rangle,
\end{split}
\tag{P3}\\
\begin{split}
Pathway_{4}:&a^{\dag}_{0}|\emptyset \mathcal{G} \rangle\xrightarrow[a_{0}|s_{1}\rangle\langle g_{1}|]{} |s_{1}\rangle\langle g_{1}|\emptyset \mathcal{G}\rangle \xrightarrow[|e_{1}\rangle\langle s_{1}|]{e^{i\phi}} |e_{1}\rangle\langle g_{1}||\emptyset \mathcal{G}\rangle\xrightarrow[c^{\dag}_{a}|g_{1}\rangle\langle e_{1}|]{} c^{\dag}_{a}|\emptyset \mathcal{G}\rangle\\
&\xrightarrow[c^{\dag}_{b}c_{a}]{}c^{\dag}_{b}|\emptyset \mathcal{G}\rangle\xrightarrow[c_{b}|e_{3}\rangle\langle g_{3}|]{} |e_{3}\rangle\langle g_{3}|\emptyset \mathcal{G}\rangle\xrightarrow[b^{\dag}_{0}|g_{3}\rangle\langle e_{3}|]{} b^{\dag}_{0}|\emptyset \mathcal{G}\rangle\xrightarrow[b^{\dag}_{l}b_{0}]{} b^{\dag}_{l}|\emptyset \mathcal{G}\rangle,
\end{split}
\tag{P4}
\end{flalign}
where $|\emptyset\rangle$ means the vacuum state in any cavities and $|\mathcal{G}\rangle$ is abbreviated form of $g_{1}g_{2}g_{3}g_{4}$ which means all the atoms are in the ground state. The arrows indicate that the states on the left go from one interaction to the states on the right. All of the interactions in the expression can be found in the Hamiltonian that describes the interactions, but for simplicity we have omitted the interaction strength and Rabi frequency from the expression. Due to the classical field, there is a phase difference $\phi$ between the two pathways, and the total probability is zero when the phase difference is $\pi$, so the coherence appears in the form of extinction interference. It can also be explained by formula, we can simply write $C(r^{b})$ to the form $A(1+e^{i\phi})$. From the formula we find that when $\phi=\pi$, $R^{b}$ can only be equal to 0. For the outgoing photon from the right port of the CRW-b, the wave function can be expressed as the unnormalized form $C(r^{b})\Psi(r^{b})=\sum_{n}u_{n}(r^{b})\psi_{n}(r^{b})$. We take the wave function corresponding to $Pathway_{1}$ to be $u_{1}(r^{b})\psi_{1}(r^{b})$, and take the wave function corresponding to $Pathway_{2}$ be $u_{2}(r^{b})e^{i\phi}\psi_{2}(r^{b})$. Under the condition of $\Omega_{1}=\Omega_{2}$, by comparing expressions $A(1+e^{i\phi})\Psi(r^{b})$ and $u_{1}(r^{b})\psi_{1}(r^{b})+u_{2}(r^{b})e^{i\phi}\psi_{2}(r^{b})$, $u_{1}(r^{b})\psi_{1}(r^{b})=u_{2}(r^{b})\psi_{2}(r^{b})$ can be obtained. Since $\psi_{1}(r^{b})$ and $\psi_{2}(r^{b})$ have different state vectors, only the state vectors shared by aa and bb have coefficients that are nonzero. This indicates that when the Rabi frequency of the two classical fields is the same, the probability of the photon transport in the two pathways is equal.

In Figs. 4(a) and 4(b) we plot the $R^{a}$ spectra as a function of phase difference when the detuning is equal to zero, and the spectra as a function of detuning when the phase difference is $\pi$, respectively. By comparing the two figures, we find that the probability of the resonant photon output from the right port of the CRW-a is zero when the phase difference is $\pi$. By simplifying Eq. (A2) describing the amplitude $r^{a}$ as before, we can obtain the form $A(X_{1}X_{3}+e^{i\phi}X_{2}^{2})+C_{0}/C'_{0}$, where $C_{0}$ and $C'_{0}$ are expressed as before. It can be obtained by calculation that when $k$ approaches $\pi$, the value of $X_{1}X_{3}$ and $X_{2}^{2}$ are equal. Therefore $r^{a}$ can be obtained to be zero when the phase difference is $\pi$ and the frequency resonance are reached.

\begin{figure}[ht]
  \centering
  \begin{subfigure}
  \centering
  \includegraphics[width=7cm,height=5.5cm]{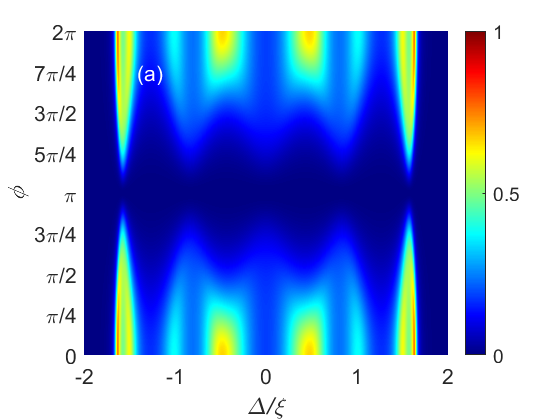}
  \end{subfigure}
    \centering
  \begin{subfigure}
  \centering
  \includegraphics[width=7cm,height=5.5cm]{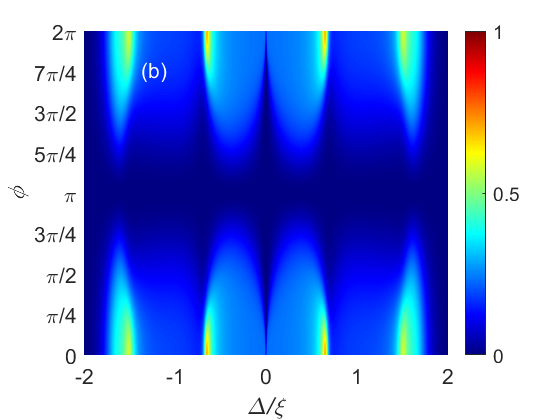}
  \end{subfigure}
    \centering
  \caption{(Color online) $R^{b}$ as a function of $\Delta$ and $\phi$. (a) $l$=1. (b) $l$=2. When the phase difference is $\pi$, $R^{b}$ is constant zero. For convenience, the other parameters are $\xi=1$, $g_{a}=g_{s}=0.5$, $\Omega_{1}=\Omega_{2}=0.5$, $\omega=10$. }
\end{figure}

When the photon exits from the left port of the CRW-a, the pathway of photon transport in the router is different from that of the other three ports. By comparing Fig. 5(a) and 5(b), neither $R^{a}$ nor $L^{b}$ is symmetric with respect to $\phi=\pi$ when the energy $E$ of the incident photon does not resonate with the frequency $\omega$. Under the condition that $\Omega_{1}=\Omega_{2}$, $R^{b}$ can also be symmetric with respect to $\phi=\pi$. $L^{a}$ is symmetric whether or not the condition $\Omega_{1}=\Omega_{2}$ exists. It indicates that the coherent pathways of the outgoing photon from the left port of CRW-a is not affected by the phase difference as the other three routing rates. As can easily be seen from the scheme in Fig. 1, pathways can be divided into two types according to whether the wave function has a phase or not. Photons in one type of these pathways return along the pathway they occurred in. If the photon is coupled to Atom 1, the pathway will experience both $|s_{1}\rangle\stackrel{\Omega_{1}e^{i\phi}}{\longrightarrow}|e_{1}\rangle$ and $|e_{1}\rangle\stackrel{\Omega_{1}^{*}e^{-i\phi}}{\longrightarrow}|s_{1}\rangle$ transitions, and if the photon is not coupled to Atom 1, it will not experience a transition with phase regulation. But the wave functions corresponding to these paths have no phase whether they are coupled to Atom 1 or not. Thus the interference between pathways is not phase regulated. In the other type, the pathway of the photon will traverse all four node cavities in the router in a clockwise or counterclockwise direction and return to the $a_{0}$ cavity. In the pathway where the direction is clockwise, the photon will undergo $|s_{1}\rangle\stackrel{\Omega_{1}e^{i\phi}}{\longrightarrow}|e_{1}\rangle$, while in the pathway counterclockwise, the phooton will undergo $|e_{1}\rangle\stackrel{\Omega_{1}^{*}e^{-i\phi}}{\longrightarrow}|s_{1}\rangle$ transitions, respectively. So we can write the total unnormalized wave function as $C(l^{a})\Psi(l^{a})=\sum_{n}u_{n}(l^{a})\psi_{n}(l^{a})+u_{c}(l^{a})e^{i\phi}\psi_{c}(l^{a})+u_{cc}(l^{a})e^{-i\phi}\psi_{cc}(l^{a})$, where $\psi_{n}(l^{a})$ mean the wave functions of the pathways in type one, $\psi_{c}(l^{a})$ and $\psi_{cc}(l^{a})$ is the pathway clockwise and counterclockwise, respectively. Reducing $C(l^{a})$ in Eq. (A1) to the form $A+Bcos\phi$, $u_{c}(l^{a})\psi_{c}(l^{a})=u_{cc}(l^{a})\psi_{cc}(l^{a})$ can be obtained, And since aa and bb are composed of the same state vector, the two paths have the same wave function under this condition. This means that the probability of a photon following a pathway that traverses the four node cavities clockwise or counterclockwise is the same. We therefore conclude that the probability of these paths is the same when the photons couple the same number of atoms and cavities regardless of the order in different pathways, and other parameters such as Rabi frequency are the same.

\begin{figure}[ht]
  \centering
  \begin{subfigure}
  \centering
  \includegraphics[width=7cm,height=5.5cm]{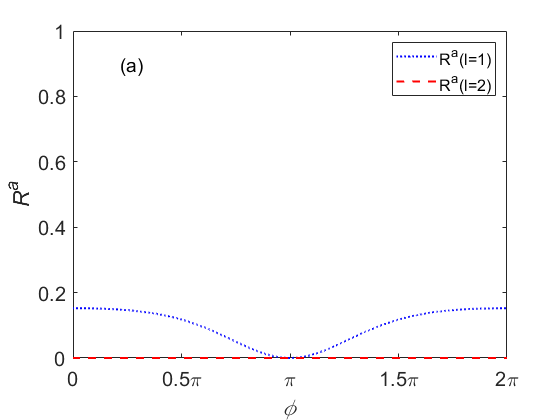}
  \end{subfigure}
    \centering
  \begin{subfigure}
  \centering
  \includegraphics[width=7cm,height=5.5cm]{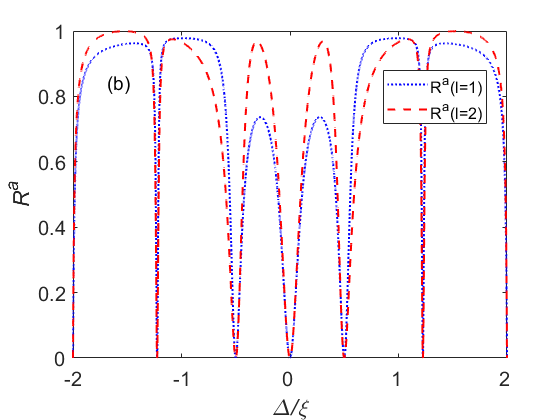}
  \end{subfigure}
    \centering
  \caption{(Color online) (a) $R^{a}$ as a function of $\phi$ with $\Delta=0$. (b) $R^{a}$ as a function of $\Delta$ with $\phi=\pi$. At conditions $\Delta=0$ and $\phi=\pi$ both reached, $R^{a}$ is zero and independent of $l$. For convenience, the other parameters are $\xi=1$, $g_{a}=g_{s}=0.5$, $\Omega_{1}=\Omega_{2}=0.5$, $\omega=10$. }
\end{figure}

\begin{figure}[ht]
  \centering
  \begin{subfigure}
  \centering
  \includegraphics[width=7cm,height=5.5cm]{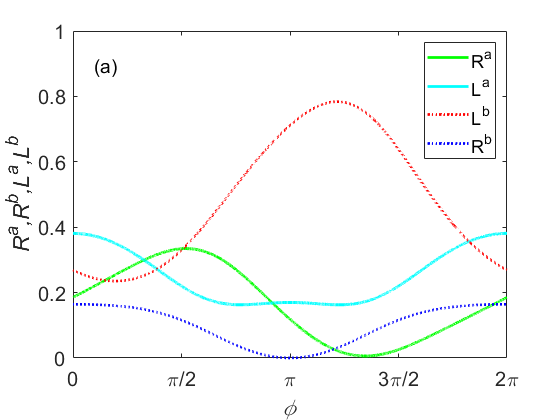}
  \end{subfigure}
    \centering
  \begin{subfigure}
  \centering
  \includegraphics[width=7cm,height=5.5cm]{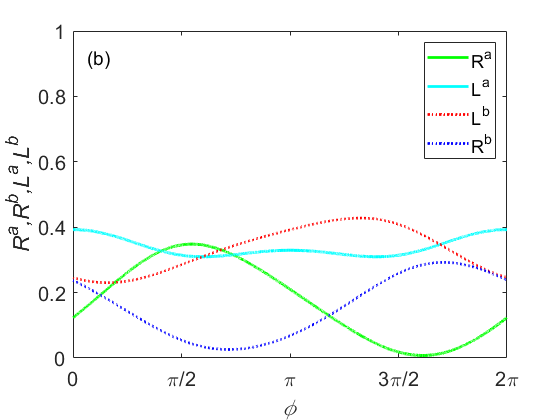}
  \end{subfigure}
    \centering
  \caption{(Color online) $L^{a}(E)$ (solid cyan line), $L^{b}(E)$ (dashed red line), $R^{a}(E)$ (solid green line) and $R^{b}(E)$ (dashed blue line) as a function of the energy of the incident wave. (a) $\Omega_{2}=0.5$, $L^{a}$ and $R^{b}$ are symmetric with respect to phase $\phi=\pi$. (b) $\Omega_{2}=0.7$, only $L^{a}$ is symmetric with respect to phase $\phi=\pi$. For convenience, the other parameters are $\xi=1$, $g_{a}=g_{s}=0.65$, $\Omega_{1}=0.5$, $E=10.5$, $\omega=10$, $l=1$. }
\end{figure}

We next investigate the effect of the classical field and the coupling strength on the scattering frequency. Since the Fig. 2 is symmetric with respect to the resonance frequency, we will next plot only the part of $\Delta>0$ for simplicity. It can be seen from Eqs. (9b)-(9e) that the values of $\omega_{+}(\omega'_{+})$ and $\omega_{-}(\omega'_{-})$ are symmetric with respect to $E=\omega$, so only $\omega_{+}$ and $\omega'_{+}$ are discussed. Figure 6 shows the scattering frequency as a function of $g_{s}$ and $\Omega$. Where the dashed line represents $\omega_{+}$ and the solid line represents $\omega'_{+}$. The red line represents $g_{s}=0.2$, while blue and green denote $g_{s}=0.5$ and $g_{s}=0.8$, respectively. From the figure, it can be found that both $\omega_{+}$ and $\omega'_{+}$ shift in the direction of increasing detuning as $\Omega$ increases. As $g_{s}$ increases, $\omega_{+}$ and $\omega'_{+}$ gradually move closer to each other and as a whole shift towards decreasing detuning.

 \begin{figure}[ht]
  \centering
  \includegraphics[width=7cm,height=5.5cm]{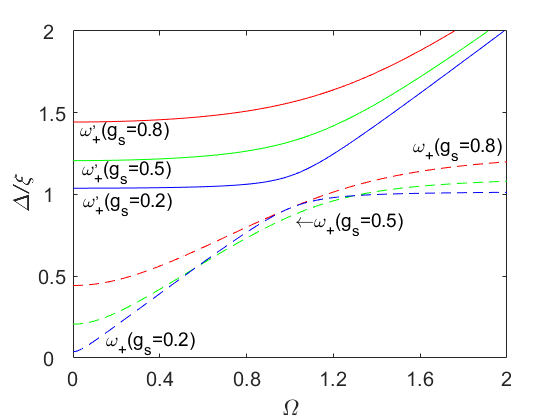}
    \centering
  \caption{(Color online) $\omega_{+}$ (solid red line, solid green line and solid blue line) and $\omega'_{+}$ (dashed red line, dashed green line and dashed blue line) as a function of Rabi frequency $\Omega$ and coupling strength $g_{s}$. For convenience, the other parameters are $\xi=1$, $\omega=10$. }
\end{figure}

 In the previous investigations, we have considered the scattering frequency as the frequency at which the router as a whole scattered the photons, as shown in Fig. 7(a). In Figure 7(a), we set $\Omega_{1}=\Omega_{2}$, $\omega_{+1}=\omega_{+2}$ and $\omega'_{+1}=\omega'_{+2}$ can be obtained. Therefore, there are two scattering frequencies in the figure, which are indicated by the black dashed line and the dotted line, respectively. When the frequency of the incident photon is not equal to the scattering frequency, the photon tends to transport freely in the CRW-a and transmit out from the right port. Under the condition that the frequency of the incident photon is resonant with the scattering frequency, the photon can exit from the left port of CRW-a and CRW-b. Figures 7(b) and 7(c) show the $L^{a}$ spectra for smaller and larger $|\Delta_{\Omega}|=|\Omega_{1}-\Omega_{2}|$, respectively. In Fig. 7(b), it can be seen that when $|\Delta_{\Omega}|$ is small, the position of the peak is near the two scattering frequencies, reflected by the merging of the two peaks into a single peak. While when $|\Delta_{\Omega}|$ is large, each scattering frequency corresponds to a peak, as shown by the blue line in Fig. 7(c). When the peak height corresponding to one of the scattering frequencies is not enough, it will merge with the other peak, and the result is that only one peak is presented, as shown by the red line in Fig. 7(c). Thus we consider the total routing probability as the direct sum of the routing probabilities corresponding to each scattering frequency. This is also the reason why the frequency of the peak tip will deviate from the scattering frequency under some conditions, such as $g_{a}$ is larger, as shown in Fig. 7(d).

\begin{figure}[ht]
  \centering
  \begin{subfigure}
  \centering
  \includegraphics[width=7cm,height=5.5cm]{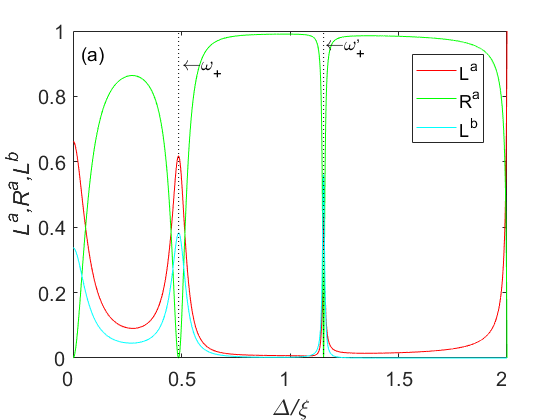}
  \end{subfigure}
    \centering
  \begin{subfigure}
  \centering
  \includegraphics[width=7cm,height=5.5cm]{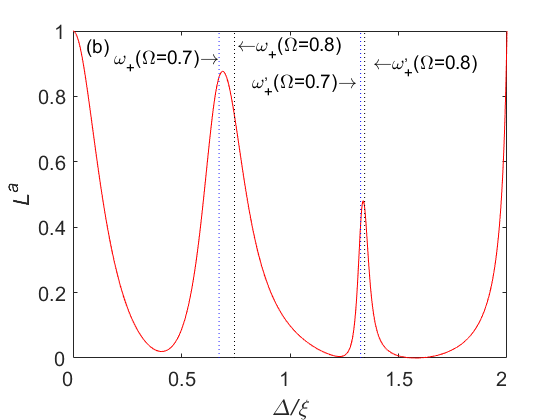}
  \end{subfigure}
    \centering
  \begin{subfigure}
  \centering
  \includegraphics[width=7cm,height=5.5cm]{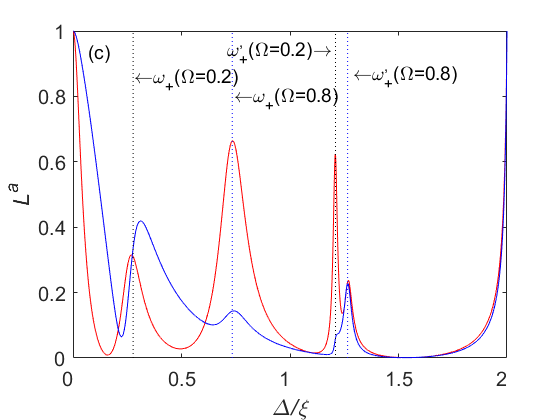}
  \end{subfigure}
    \centering
  \begin{subfigure}
  \centering
  \includegraphics[width=7cm,height=5.5cm]{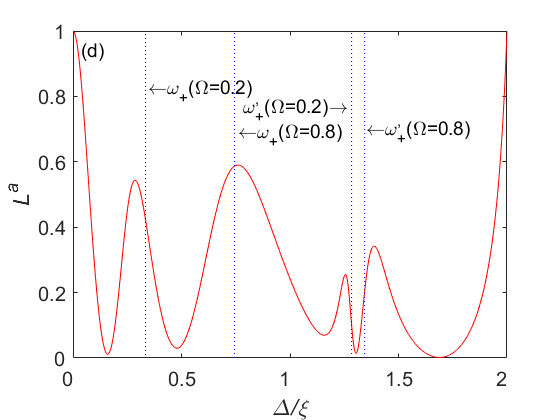}
  \end{subfigure}
  \caption{(Color online) $L^{a}$ (solid red line and solid blue line), $L^{b}$ (solid cyan line) and $R^{a}$ (solid green line) as a function of energy $E$. (a) $\Omega_{1}=\Omega_{2}=0.5$, $g_{a}=g_{s}=0.4$. The frequency of the peak is resonant with the scattering frequency. (b) $\Omega_{1}=0.7$, $\Omega_{2}=0.8$, $g_{a}=g_{s}=0.6$. The spectra of $L^{b}$ has only two peaks, and the peaks do not resonate with the scattering frequency. (c) $g_{a}=g_{s}=0.5$. $\Omega_{1}=0.8$ and $\Omega_{2}=0.2$ for red line, $\Omega_{1}=0.2$ and $\Omega_{2}=0.8$ for blue line. The four peaks of the red line correspond to one scattering frequency each. The peak corresponding to the scattering frequency $\omega'_{+}$ at $\Omega=0.2$ in the blue line has a lower height and merges into a single peak with the peak corresponding to $\omega'_{+}$ at $\Omega=0.8$. (d) $\Omega_{1}=0.8$, $\Omega_{2}=0.2$, $g_{a}=0.8$ and $g_{s}=0.6$. The peak in the spectra deviates from the scattering frequency. For convenience, the other parameters are $\xi=1$, $\omega=10$ and $\phi=\pi$. }
\end{figure}

Through previous studies, we found that directional routing can be achieved in this model by adjusting position $l$, coupling strength $g_{a}$ and $g_{s}$, Rabi frequency $\Omega_{1}$ and $\Omega_{2}$, and phase difference $\phi$, as shown in Fig. 8(a) and 8(b). In Fig. 8(a), the photon with incident frequency near $\omega_{+}$ exits from the right port of CRW-b with probability greater than 0.95 and from the other three ports with probability almost zero.  In Fig. 8(b), photons resonant with the cavity mode and near resonant with $\omega_{+}$ have a probability greater than 0.95 to exit from the left side of the CRW-b. Photons directed from the left port and right port of CRW-a are better modulated, as can be seen from FIG. 2(a). Therefore, we believe that the model can play a helpful role in single-photon directional routing.

\begin{figure}[ht]
  \centering
  \begin{subfigure}
  \centering
  \includegraphics[width=7cm,height=5.5cm]{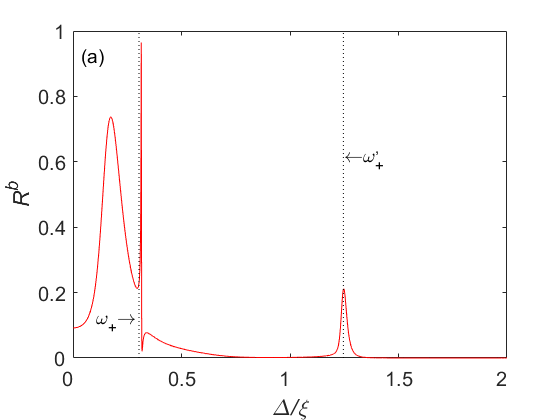}
  \end{subfigure}
    \centering
  \begin{subfigure}
  \centering
  \includegraphics[width=7cm,height=5.5cm]{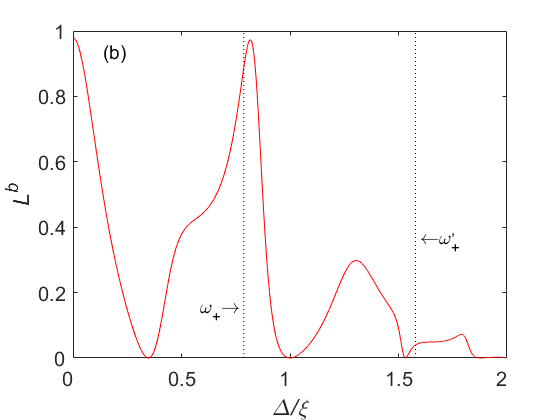}
  \end{subfigure}
    \centering
  \caption{(Color online) (a) $R^{b}$ as a function of energy $E$. $\Omega_{1}=\Omega_{2}=0.2$, $g_{a}=0.58$, $g_{s}=0.55$ and $\phi=0$. There is a steep peak with height greater than 0.95 near $E=\omega_{+}$. (b) $\Omega_{1}=\Omega_{2}=0.7$, $g_{a}=1$, $g_{s}=0.9$ and $\phi=\pi$. There are two peaks with height greater than 0.95 where the detuning is 0 and $E=\omega_{+}$. For convenience, the other parameters are $\xi=1$, $\omega=10$ and $l=9$. }
\end{figure}

\section{Controlling non-reciprocal directions based on phase}

In the previous section, we set the coupling strengths and the frequencies of atomic levels to be the same, respectively. In the following, we set an energy difference $\Delta_{es1(4)}=\omega_{e1(4)}-\omega_{s1(4)}$ between the frequencies of the intermediate state and the excited state in the two three-level atoms, and regulate each coupling strength independently.

When two detunings $\Delta_{es1(4)}$ are nonzero, the spectra of the four routing probabilities become asymmetric. In Fig. 9 we plot the probabilities $L^{b}$ as a function of $\Delta_{es1}$ and $\Delta_{es4}$, respectively. It can be found that the both $\Delta_{es1}$ and $\Delta_{es4}$ can cause the peaks split and energy shift. In addition, it will also affect the routing probability.

\begin{figure}[ht]
  \centering
  \begin{subfigure}
  \centering
  \includegraphics[width=7cm,height=5.5cm]{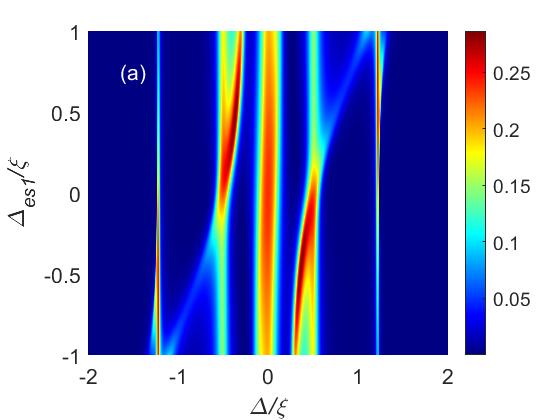}
  \end{subfigure}
    \centering
  \begin{subfigure}
  \centering
  \includegraphics[width=7cm,height=5.5cm]{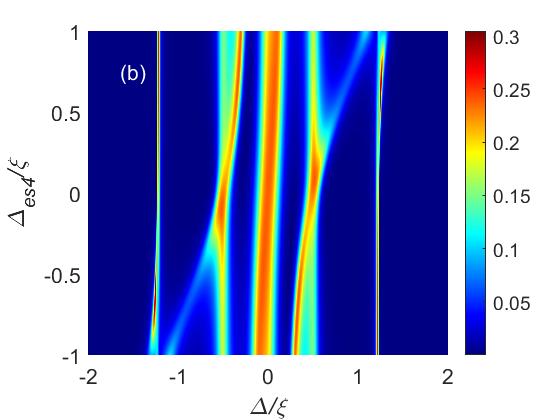}
  \end{subfigure}
    \centering
  \caption{(Color online) (a) $L^{b}$ as a function of energy $E$ and detuning $\Delta_{es1}$. (b) $L^{b}$ as a function of energy $E$ and detuning $\Delta_{es4}$. $\Omega_{1}=\Omega_{2}=g_{a}=g_{s}=0.5$ and $\phi=0$. As the absolute value of $\Delta_{es1(2)}$ increases, the peak produces splitting and energy shift, and the height of the peak changes accordingly. The other parameters are $\xi=1$, $\omega=10$ and $l=1$. }
\end{figure}

By rewriting the terms in Eq. 10 as $e^{ik_{a}j}+l^{a}e^{-ik_{a}j}\rightarrow l^{a}e^{-ik_{a}j}$ and $l^{b}e^{-ik_{b}j}\rightarrow e^{ik_{a}j}+l^{b}e^{-ik_{b}j}$ and recalculating, the probability of photon transmission from the left side of CRW-b to the left side of CRW-a can be obtained and expressed in $T_{l^{b}}$. We define the difference $N_{l^{b}l^{a}}=L^{b}-T_{l^{b}}$ between $L^{b}$ and $T_{l^{b}}$ to describe the nonreciprocity of photon transmission between these two ports. When $N_{l^{b}l^{a}}\rightarrow1$, the photon incident from the left side of CRW-a can be transmitted to the left side of CRW-b almost perfectly, but the probability of the photon incident from the left side of CRW-b exiting from the left side of CRW-a is approximately 0. Therefore, we consider that nonreciprocal photon transmission from the left side of CRW-a to the left side of CRW-b is achieved when $N_{l^{b}l^{a}}\rightarrow1$, and nonreciprocal photon transmission in the opposite direction is achieved when $N_{l^{b}l^{a}}\rightarrow-1$.

\begin{figure}[ht]
  \centering
  \begin{subfigure}
  \centering
  \includegraphics[width=7cm,height=5.5cm]{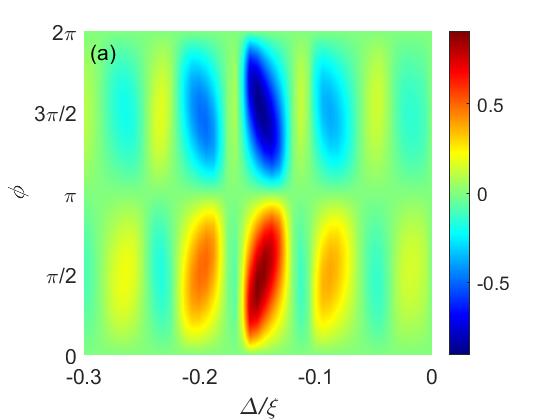}
  \end{subfigure}
    \centering
  \begin{subfigure}
  \centering
  \includegraphics[width=7cm,height=5.5cm]{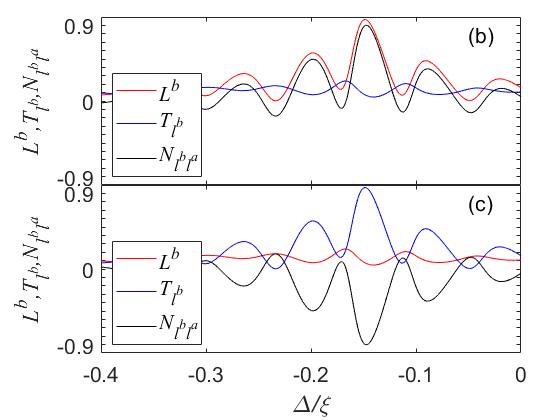}
  \end{subfigure}
    \centering
  \begin{subfigure}
  \centering
  \includegraphics[width=7cm,height=5.5cm]{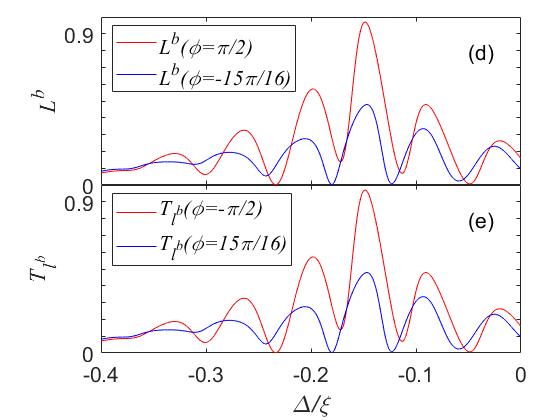}
  \end{subfigure}
    \centering
  \begin{subfigure}
  \centering
  \includegraphics[width=7cm,height=5.5cm]{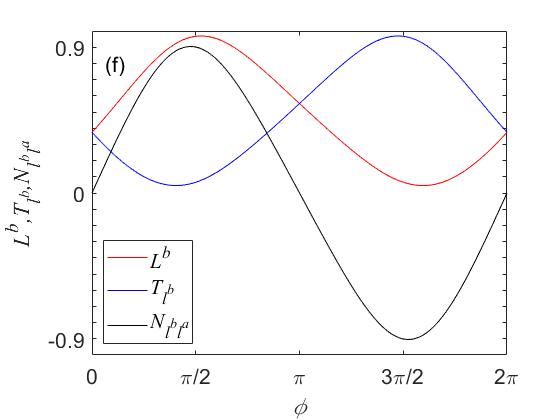}
  \end{subfigure}
  \caption{(Color online) (a) $N_{l^{b}l^{a}}$ as a function of detuning $\Delta$ and the phase difference $\phi$. The red areas represent $N_{l^{b}l^{a}}$ greater than 0 and the blue areas represent $N_{l^{b}l^{a}}$ less than 0. $L^{b}$ (red line), $T_{l^{b}}$ (blue line) and $N_{l^{b}l^{a}}$ (black line) as a function of $\Delta$ with (b) $\phi=\pi/2$ and (c) $\phi=3\pi/2$. In (b), the maximum value of the black line is greater than 0.9, and in (c), the minimum value of the black line is less than 0.9. (d) $L^{b}$ as a function of $\Delta$ with $\phi=\pi/2$ (red line) and $\phi=-15\pi/16$ (blue line). (e) $T_{l^{b}}$ as a function of $\Delta$ with $\phi=-\pi/2$ (red line) and $\phi=15\pi/16$ (blue line). (f) $L^{b}$ (red line), $T_{l^{b}}$ (blue line) and $N_{l^{b}l^{a}}$ (black line) as a function of $\Delta$ with $\Delta\sim-1.5$. As the phase changes, the values of $L^{b}$ and $T_{l^{b}}$ are flipped and $N_{l^{b}l^{a}}$ goes from greater than 0.9 to less than -0.9. The other parameters are $\xi=1$, $g_{a1}=0.57$, $g_{a2}=0.33$, $g_{b3}=0.79$, $g_{b4}=0.37$, $g_{c1}=0.98$, $g_{c3}=0.41$, $g_{d2}=0.9$, $g_{d4}=0.93$, $\Omega_{1}=0.48$, $\Omega_{1}=0.87$, $\Delta_{es1}=-\Delta_{es4}=-0.46$ and $\omega=10$. }
\end{figure}

In Fig. 10(a), we plot $N_{l^{b}l^{a}}$ as a function of detuning $\Delta$ and the phase difference $\phi$, it can be found that there is a peak of $N_{l^{b}l^{a}}$ near the phase difference $\phi=\pi/2$, while near the phase difference $\phi=-\pi/2$, there is a valley of $N_{l^{b}l^{a}}$. In Figs. 10(b) and 10(c), the spectra of $T_{l^{b}}$, $L^{b}$ and $N_{l^{b}l^{a}}$ as a function of $\Delta$ are plotted with $\phi=\pm\pi/2$, respectively. From Fig. 10(b), it can be found that when $\phi=\pi/2$, the photon transmission probability from the left side of CRW-a to the left side of CRW-b is close to 1, while the photon transmission probability in the opposite direction is close to 0, while Fig. 10(c) shows the opposite situation. This shows that in these two phases, the photon can achieve forward and reverse non-reciprocal transmission, respectively, and the probability is greater than 0.9. And we find that the red line in Fig. 10(b) is the same as the blue line in Fig. 10(c). In order to verify whether this phenomenon is related to the value of the phase, we plot Figs. 10(d) and 10(e). It can be found that the lines of the same color in Fig. 10(d) and Fig. 10(e) are exactly the same, which means $L^{b}(\phi)=T_{l^{b}}(-\phi)$. This is due to the fact that the photon experiences a transition of $|s_{1}\rangle\stackrel{\Omega_{1}e^{i\phi}}{\longrightarrow}|e_{1}\rangle$ in the pathway from the left side of CRW-a to the left side of CRW-b, and the phase of $\phi$ exists in the total wave function of the wave function superposition of all paths. However, photons transmitted in the opposite direction experience the transition of $|e_{1}\rangle\stackrel{\Omega_{1}e^{-i\phi}}{\longrightarrow}|s_{1}\rangle$, and the phase of $-\phi$ exists in the corresponding total wave function. It shows that when the input port and the output port are opposite, the probability of routing in the quantum router is affected by the phase regulation. In Fig. 10(f), we plot the spectra of $T_{l^{b}}$, $L^{b}$, and $N_{l^{b}l^{a}}$ at $\Delta\sim -1.5$ as a function of the phase difference. It is confirmed that the value of $N_{l^{b}l^{a}}$ can be changed from $1$ to $-1$ by the phase. Therefore, not only realize the non-reciprocal transmission, but also the reversal of the non-reciprocal port of a single photon can be realized by regulating the phase separately.

\section{conclusions}

In summary, we have proposed a single photon quantum router model consisting of two coupled-resonator waveguides and four single cavities, and discuss the effect of phase difference and other parameters on the single photon routing in the model. In this model, we increase the number of nodal cavities to realize the existence of multiple pathways between the incident photon and the outgoing port. Due to the interference between pathways, probabilistic waves can realize destructive interference or constructive interference, so as to realize channel closing or directional routing. From the perspectives of analytical solution and model, we have explained the reason why the probability of different routes is zero respectively. Under the premise that the two Rabi frequencies are equal, when the phase difference is set as $\pi$, photons of any frequency in the energy band output from the right port of the CRW-b are prohibited. For the right port of CRW-a, only resonant photons are prohibited from emitting. In addition, because the reflected photon may traverse the four nodal cavities clockwise or counterclockwise, it is affected by the phase difference differently from the photon emitted from the other three ports. We also study the influence of parameters on the scattering frequency of the router and the reason why the peak in the spectra deviates from the scattering frequency. The first case, because the difference between the two Rabi frequencies is small, the corresponding scattering frequency difference is small, and the summit deviates from the scattering frequency. The second case, due to the increase of the coupling strength between the atom and the vertical cavity, the summit of the spectral line deviates from the scattering frequency, even though another coupling strength does not affect the scattering frequency. By adjusting the parameters, we successfully get the output probability of four ports close to $100\%$, which successfully realizes the directional routing. In addition, we also find the relationship between the wave function of the photon transmitted in the opposite direction and the wave function of the original photon. The non-reciprocal transmission of photons is realized, and the direction of non-reciprocal transmission is successfully changed by adjusting the phase without changing the frequency of incident photons. This is an effective method that can be designed experimentally, and we expect these results to be helpful for the research on multi-channel quantum routers.

\section{Appendix}
\begin{flalign}
\begin{split}
l^{a}&=(2i\xi V_{1}sink(-(-1+e^{2ikl})^{2}(X_{1}X_{3}-X_{4}^{2})X_{5}+2\xi sink(-i(-1+e^{2ikl})(V_{1}(X_{1}X_{3}-X_{4}^{2})+V_{2}(X_{1}+X_{3})X_{5})\\
&+2\xi V_{2}sink(V_{1}(X_{1}+X_{3}-e^{2ikl}X_{3})+V_{2}X_{5}-2i\xi V_{1}V_{2}sink))))/G-1,
\end{split}
\label{eqA.1}
\tag{A1}\\
r^{a}&=(4 i \xi^2 V_{1} V_{2} sin^{2}k (-i e^{-i \phi} (-1 + e^{2 i k l}) (X_{2} X_{4} + e^{i \phi} X_{1} X_{5}) +2 \xi sink (V_{1} X_{1} + V_{2} X_{5} - 2 i \xi V_{1} V_{2} sink)))/G,
\label{eqA.2}
\tag{A2}\\
\begin{split}
l^{b}&=-((2 i e^{-i \phi} \xi V_{1} sink ((-1 + e^{2 i k l})^2 X_{2} (X_{1} X_{3} - X_{4}^{2}) -2 \xi V_{2} sink (-i (-1 + e^{2 i k l}) X_{2} (X_{1} + X_{3}) +2 \xi (V_{2} X_{2} \\
&+e^{i (2 k l + \phi)} V_{1} X_{4}) sink)))/G,
\end{split}
\label{eqA.3}
\tag{A3}\\
r^{b}&=(4 i e^{-i \phi}\xi^{2} V_{1} V_{2} sin^{2}k (-i (-1 + e^{2 i k l}) (X_{2} X_{3} + e^{i \phi} X_{4} X_{5}) +2 \xi (V_{2} X_{2} + e^{i \phi} V_{1} X_{4}) sink))/G,
\label{eqA.4}
\tag{A4}\\
\begin{split}
G&=((-1 + e^{2 i k l})^{2} (X_{1} X_{3} - X_{4}^{2}) (-X_{2}^{2} + X_{1} X_{5}) +2 i (-1 + e^{2 i k l}) \xi (V_{1} (X_{1} X_{3} - X_{4}^{2}) (X_{1} + X_{5}) +V_{2} (X_{1} + X_{3})\\
 &(-X_{2}^{2} + X_{1} X_{5})) sink -4 \xi^{2} (V_{1}^{2} (X_{1} X_{3} - X_{4}^{2}) + V_{1} V_{2} (X_{1} + X_{3}) (X_{1} + X_{5}) +V_{2}^{2} (-X_{2}^{2} + X_{1} X_{5}) -e^{2 i k l} V_{1} V_{2}\\
  &(X_{1} (X_{3} + X_{5}) + 2 X_{2} X_{4} cos\phi)) sin^{2}k +8 i \xi^{3} V_{1} V_{2} (V_{1} (X_{1} + X_{3}) + V_{2} (X_{1} + X_{5})) sin^{3}k +16 \xi^{4} V_{1}^{2} V_{2}^{2} sin^{4}k).
\end{split}
\label{eqA.5}
\tag{A5}
\end{flalign}

%\begin{thebibliography}{99}
%\end{thebibliography}
\bibliography{articlenew}
\end{document}